\documentclass[aps,pre,floatfix,superscriptaddress,twocolumn,nofootinbib]{revtex4}

\usepackage[dvips]{graphicx}
\usepackage{amssymb,amsfonts,amsmath}

\let\a=\alpha \let\b=\beta     \let\e=\varepsilon
     \let\th=\theta  
        \let\x=\xi         \let\r=\rho
\let\s=\sigma \let\t=\tau    
   
 \let\D=\Delta   
         
 \let\Y=\Upsilon 

\let\io=\infty
\def\ie{{i.e. }}\def\eg{{e.g. }}

\def\FF{{\cal F}} 
\def\NN{{\cal N}}

\def\to{\rightarrow}

\newcommand{\beq}{\begin{equation}}
\newcommand{\eeq}{\end{equation}}

\begin{document}

\title{
Dynamically correlated regions and configurational entropy in supercooled liquids
}

\author{Simone Capaccioli}
\affiliation{Dipartimento di Fisica, Universit\`a di Pisa, Largo B. Pontecorvo
3, 56127, Pisa, Italy}
\affiliation{CNR-INFM/CRS-Soft, Universit\`a di Roma ``La Sapienza'',
P.le A. Moro 2, 00185 Roma, Italy}

\author{Giancarlo Ruocco}
\affiliation{CNR-INFM/CRS-Soft, Universit\`a di Roma ``La Sapienza'',
P.le A. Moro 2, 00185 Roma, Italy}
\affiliation{Dipartimento di Fisica, Universit\`a di Roma ``La Sapienza'',
P.le A. Moro 2, 00185 Roma, Italy}

\author{Francesco Zamponi}
\affiliation{Service de Physique Th\'eorique, DSM/CEA/Saclay, 91191 Gif-sur-Yvette
Cedex, France}
\affiliation{Laboratoire de Physique Th\'eorique de l'\'Ecole Normale Sup\'erieure,
24 Rue Lhomond, 75231 Paris Cedex 05, France}

\begin{abstract}
When a liquid is cooled below its melting temperature, if
crystallization is avoided, it forms a glass. This phenomenon,
called {\it glass transition}, is characterized by a marked
increase of viscosity, about 14 orders of magnitude, in a narrow
temperature interval. The microscopic mechanism behind the glass
transition is still poorly understood. However, recently, great
advances have been made in the identification of {\it cooperative
rearranging regions}, or {\it dynamical heterogeneities}, \ie
domains of the liquid whose relaxation is highly correlated. The
growth of the size of these domains is now believed to be the
driving mechanism for the increase of the viscosity. Recently a
tool to quantify the size of these domains has been proposed. 
We apply this tool to a wide class of materials to
investigate the correlation between the size of the
heterogeneities and their configurational entropy, \ie the number
of states accessible to a correlated domain. We find that the
relaxation time of a given system, apart from a material dependent
pre-factor, is a {\it universal function} of the configurational
entropy of a correlated domain. As a consequence, we find that at
the glass transition temperature, the size of the domains and the
configurational entropy per unit volume are anti-correlated, as
originally predicted by the Adam-Gibbs theory. Finally, we use our
data to extract some exponents defined in the framework of the
{\it Random First Order Theory}, a recent quantitative theory of
the glass transition.
\end{abstract}

\maketitle

\section{Introduction}

Following the seminal paper of Adam and Gibbs
\cite{AG65}, the concept of {\it cooperative rearranging regions}
(CRR) has become ubiquitous in the literature on the glass
transition. In fact, the cooperative relaxation of such regions is
proposed by many theories to be the elementary rearrangement
mechanism taking place in the liquid close to the glass transition
temperature $T_g$, and the increase of the size of these regions
(or {\it dynamical correlation length} $\xi$) on decreasing the
temperature is proposed to be responsible for the dramatic slowing
down of the dynamics around~$T_g$.

Roughly speaking, the Adam-Gibbs theory proposes the existence of
CRR of size $\xi$, whose relaxation time $\tau(\xi)$ is given by:
\beq \label{tauAG} \tau(\xi) \sim \t_0 e^{ \left(
\frac{\xi}{\xi_o} \right )^d \frac{\D}{K_B T}} \, \eeq
\ie it is activated with a barrier proportional to the number of
units belonging to the CRR (here $d$ is the space dimensionality,
$K_B$ is the Boltzmann constant, $\Delta$ and $\xi_o$ are system
dependent characteristic energy- and length-scale respectively).
If we indicate by $S_c(T)$ the configurational entropy per
molecule (\ie the difference between the entropy of the liquid and
its crystal), then the logarithm of the number of accessible
states in a CRR, called $\sigma_{_{CRR}}$, a central quantity in
the AG theory, turns out to be
\beq \label{sCRRAG}
\sigma_{_{CRR}}(\xi) = \frac{S_{c}(T)}{K_B} \r \xi^d \ , 
\eeq
where $\r$ is the number density of molecules.
For a region to be able to relax, the number of accessible states
must be larger than a given threshold, let us say
$n_o$. Therefore $\sigma_{_{CRR}}(\xi) > \ln (n_o)$, which
implies by Eq.~(\ref{sCRRAG}) the existence of a lower cutoff $\tilde \xi$ on the size
of the CRR given by $\tilde \xi^d \sim K_B \ln ( n_o) /
\r S_{c}(T)$. Assuming that the dynamical decorrelation
in the liquid is dominated by the shortest relaxation time
(smaller CRR), and substituting the previous equation in
(\ref{tauAG}), one is left with the celebrated Adam-Gibbs relation
\beq \label{AG} \t(T) \sim \t_0 e^{\frac{\D \ln(n_o)}{\r\xi_o^d T
S_c(T)}} = \t_0 e^{\frac{\mu}{T S_c(T)}} \ . \eeq
As $S_c(T)$ is experimentally observed to be a decreasing function
of temperature, and seems to vanish linearly at the Kauzmann
temperature $T_K$, both $\tilde\xi$ and $\t(T)$ are predicted to
grow and diverge at $T_K$ by the AG theory. Note that -by
construction-, according to AG theory, the "entropy" of the
smallest CRR, that dominate the relaxation, is $K_B
\sigma_{_{CRR}}(T) = S_c(T) \, \r \tilde \xi(T)^d \sim K_B \ln (n_o)$,
\ie it is a temperature independent quantity.

The AG theory had an enormous impact, and the relation (\ref{AG})
has been shown to be fairly well compatible with experimental
data. Still, from the theoretical point of view, the AG scenario
is not firmly established~\cite{BB04}, mainly because of both the
unnatural scaling of $\t$ with $\x$ in Eq.~(\ref{tauAG}) (one
would naturally expect an exponent $\psi < d$ related to
the shape of the CRR-CRR interface), and, more important, the predicted
values of $\xi$ which turn out to be unreasonably small.

In the last decade, two major advances have been made in
understanding the relaxation phenomena that are behind the
Adam-Gibbs picture. Firstly, Kirkpatrick, Thirumalai and
Wolynes~\cite{KTW87} identified a deep analogy between the
behavior of supercooled liquids and that of a class of mean field
spin glass models. These models are characterized by the existence
of an exponentially large (in the system volume $V$) number $\NN$
of metastable states at low temperature, that give a finite
contribution $S_c(T) = \frac{K_B}{\r V} \ln \NN$ to the liquid
entropy, to be identified with the configurational entropy of AG.
The mean field scenario was then used as a starting point for a
nucleation theory of supercooled liquids, the {\it Random First
Order Theory} (RFOT)~\cite{XW01,LUB03,BB04}. Secondly, a method to
estimate the size of the CRR in experiments was proposed by
Berthier at al.~\cite{Be06,Da07}. We will now briefly review these
results.

As in the AG theory, in the RFOT theory the liquid close to $T_g$
is supposed to be a ``mosaic state'' made of CRR of typical radius
$\xi$; one assumes that inside a CRR the system behaves almost as
a mean field system. Then one can show~\cite{BB04} that,
for a CRR in a state $\cal{A}$, the free
energy cost for nucleation of {\it any possible state}
$\cal{B} \neq \cal{A}$ in a droplet of linear dimension $r$ is
\beq \label{barriera} \D F_{_{\cal{A}\cal{B}}}(r) = - T S_c(T)
\r r^d + \Y r^\th \ . \eeq
The configurational entropy turns out to be the driving force for
nucleation, while $\Y$ is a surface tension which is assumed to be
roughly constant around $T_g$ and the exponent $\th < d$.
Note that the free energy barrier for a given state is given only by the surface
term. The bulk term comes from the fact that the number of possible
different states is exponentially large in the volume of the droplet, \ie
there is an entropic gain in changing state inside the droplet.
The typical size $\tilde \x$ of the CRR is given by the condition $ \D
F_{_{\cal{A}\cal{B}}}(\tilde \xi)=0$, because for larger sizes
nucleation inside a CRR is not avoidable and the CRR looses its
identity~\cite{BB04}. This gives
\beq 
\label{xiRFOT} 
\tilde \xi = \left( \frac{\Y}{\r \, T  S_c}
\right)^{\frac1{d-\th}} \ . \eeq

The {\it thermodynamic} free energy barrier for nucleation of a
different state inside a CRR is the maximum of $\D
F_{_{\cal{A}\cal{B}}}(r)$, which is found in $\tilde{r} =
\left(\frac{\th}{d}\right)^{\frac1{d-\th}} \tilde \xi$ and is
given by
\beq \label{deltaF} \D F_{_{\cal{A}\cal{B}}}(\tilde{r}) \propto
\r \tilde\x^d \; T S_c(T) =  K_B T \; \sigma_{_{CRR}}(T) \ , \eeq
\ie in the RFOT the thermodynamic barrier for nucleation is given
by the total configurational entropy of a CRR of typical size
$\tilde \xi$.
Note that in RFOT, using (\ref{xiRFOT}) and (\ref{deltaF}),
$\sigma_{_{CRR}}(T) \sim S_c(T)^{-\frac{\th}{d-\th}}$ and is
expected to diverge at $T_K$, while in AG theory it is a constant
by definition.

One of the most interesting open problems in RFOT is the relation
between this thermodynamic barrier for nucleation of a droplet
with the relaxation time of the system. Usually it is assumed
that $\t \sim e^{\D F_{_{\cal{A}\cal{B}}}(\tilde r)/K_BT} \sim
e^{\sigma_{_{CRR}}(T)}$~\cite{LUB03}. More generally one can
expect that $\t \sim e^{(\sigma_{_{CRR}}(T))^\psi}$, and the
exponents $\psi$ and $\th$ can be adjusted to recover the AG
relation (\ref{AG}) without imposing Eq.~(\ref{tauAG}). While the
exponent $\psi$ is difficult to compute analytically, estimates of
the exponent $\theta$ have been obtained by mean of instantonic
techniques~\cite{Fr05,DSW05}.

The main problem in RFOT is that the domain size $\tilde \xi$ is
not directly observable; in fact, the quest for a growing
length-scale in supercooled liquids was for long unsuccessful, as
static correlation functions such as the structure factor do not
reveal any sign of long range order setting in on approaching
$T_g$. A major theoretical advance in this direction has been
achieved in the last decade by identifying a family of
dynamic~\cite{Pa99,BerthierJCP} and static~\cite{BB04,staticC}
many-points correlation functions, inspired by the mean field
models, that define correlation lengths which are predicted to
increase fast on approaching $T_g$.

Still these correlations are not directly accessible in
experiments; nevertheless experimental evidence for a growing
number of correlated units involved in the relaxation has up to
now been obtained by different techniques, see \cite{Ed00} for a
review.

Very recently, Berthier et al. \cite{Be06,Da07} proposed a very
general but still simple and direct method to measure a ``number
of correlated units'' $N_{corr}(T)$ in glass forming systems.
They were able to
relate, by a fluctuation-dissipation-like theorem, the four-point
correlations introduced in \cite{Pa99} to an easily accessible
response function, namely the derivative of a dynamic two-point
correlation (such as, for example, the intermediate scattering
function) with respect to an external control parameter such as
temperature or density. Their result can be formulated as
follows:
\beq\label{Ncorr4-max} N_{corr,4}(T) = \frac{K_B}{\D C_p(T)}
T^2 \left\{ \max_t \chi_T(t) \right \}^2 \ , \eeq
where $\chi_T(t) = \frac{d C(t)}{d T}$ is the temperature
derivative of a suitable correlation function, and
$\D C_p$ is the configurational heat capacity
per molecule at constant pressure. 
In fact Eq.~(\ref{Ncorr4-max}) is a lower bound for $N_{corr,4}$,
but one can show~\cite{Be06,Da07,BerthierJCP} that it gives a very
good estimate of this quantity;
all the details of the derivation can be found in
\cite{Da07}. 
Moreover one can simplify the analysis by
assuming that $C(t)$ has a stretched exponential form, $C(t) =
\exp(-(t/\t_\a(T))^{\b(T)})$. Then one has
\beq\label{Ncorr4-der} N_{corr,4}(T) = \frac{K_B}{\D C_p(T)}
\frac{\b(T)^2}{e^2} \left(\frac{d\ln \t_\a}{d\ln T}\right)^2 \ ,
\eeq
plus two corrections: one involving $\frac{d\b(T)}{dT}$, the other
coming from a shift of the maximum of $\chi_T$ that -for large
stretching- is not found in $t = \t_\a$. Both corrections can be
shown to be irrelevant for the following analysis, giving an error
of the order of $1\%$ on the value of $N_{corr,4}$. Note also that
the temperature dependence of the stretching parameter $\b(T)$ is
weak around $T_g$, especially when plotted against $\ln (\t_\a)$,
see \eg Fig.~2c in \cite{Dixon90}. Experimental values of
$N_{corr,4}$ were reported in \cite{Be06,Da07} for some
prototypical glass-forming systems and it was shown that this quantity indeed
increases on approaching $T_g$.

The aim of this paper is to compute $N_{corr,4}$ for a large
number of glass forming materials and compare it to the
configurational entropy. We {\it hypothesize that $N_{corr,4}$ is
representative of the size of a CRR}, \ie that $N_{corr,4} \sim
\r \tilde \xi^d$; this identification between dynamically correlated
units and CRR is, in general, non-trivial, as it is possible to
give examples where it does not hold~\cite{Da07,BerthierJCP}.
Still, in the specific case of glass-forming liquids, it is at
least partially supported by theoretical
arguments~\cite{BB04,Da07,BerthierJCP,staticC}.

\begin{figure}[t]
\includegraphics[width=8cm]{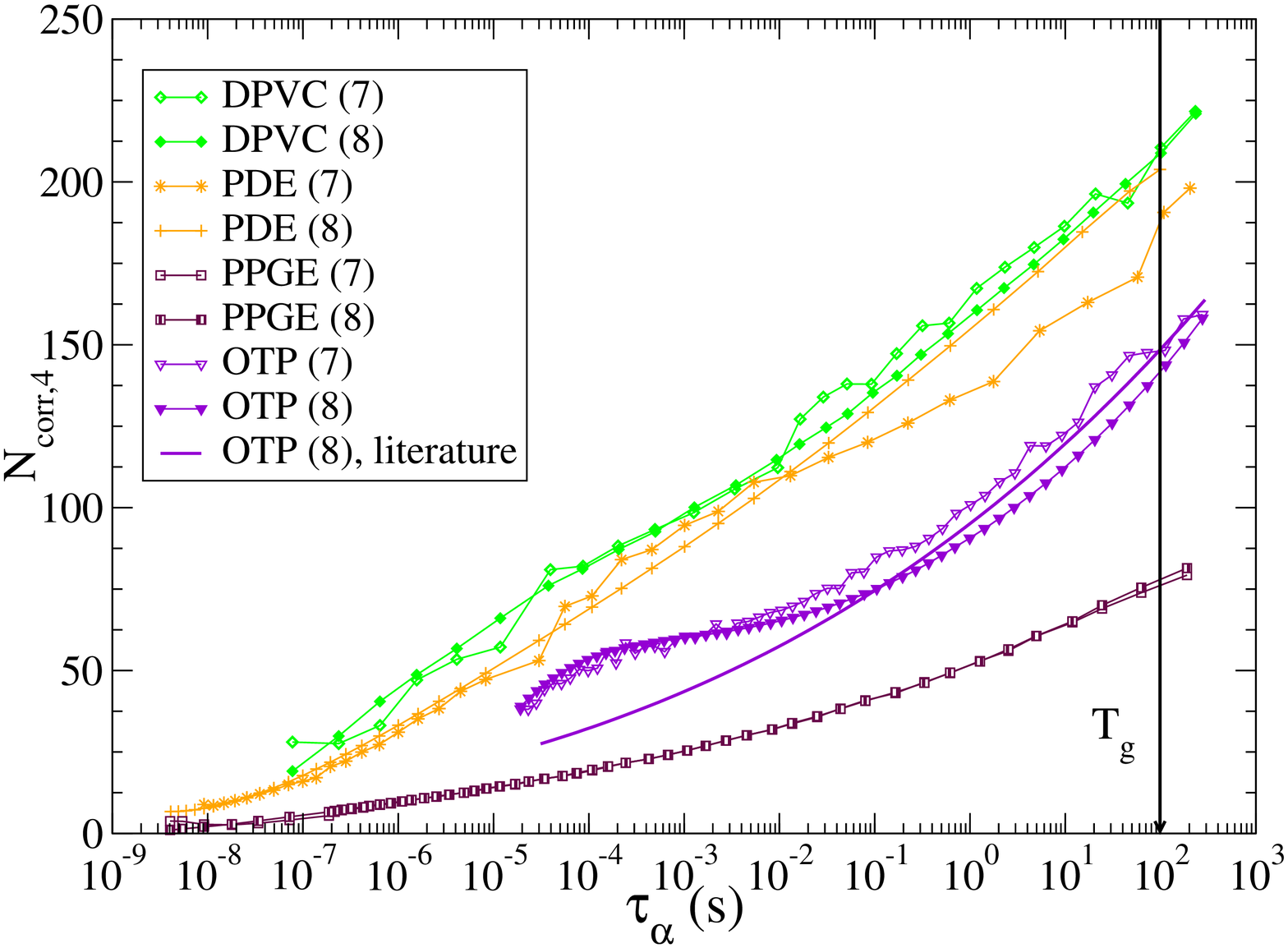}
\caption{Original data for
$N_{corr,4}$, computed by {\it i)} direct measurement of the
dielectric spectra using Eq.~(\ref{Ncorr4-max}), and {\it ii)}
fitting the measured $\t_\a(T)$ and using
Eq.~(\ref{Ncorr4-der}), for four different materials.
In the case of OTP we compare our data with the estimate of $N_{corr,4}$ obtained
from the VFT fit reported in \cite{RICH98}. A discrepancy is observed
at small times, while in the interesting region of large relaxation times
all the results are consistent.
\label{figconfronto}}
\end{figure}

Inspired by the AG and RFOT theories, following Eq.~(\ref{sCRRAG}), we define
the logarithm of available states in a correlation volume as
\beq\label{Scorr} \sigma_{_{CRR}}(T) = \frac{S_c}{K_B} N_{corr,4} = \frac{S_c}{\D C_p} \frac{\b^2}{e^2}
\left(\frac{d\ln \t_\a}{d\ln T}\right)^2 \ , \eeq
As we discussed above, AG theory predicts $\sigma_{_{CRR}}$ to be
independent of temperature, while RFOT predicts that it is an
increasing function of temperature, and predicts relations between
the relaxation time $\t_\a(T)$ and $\sigma_{_{CRR}}(T)$. The aim
of this work is to test these predictions.

From the experimental point of view, there is an important
advantage in working with $\sigma_{_{CRR}}$ instead of
$N_{corr,4}$. In fact in experiments one usually deals with
molecular liquids or polymers, where the constituents of the
liquid are complex units and it is not clear {\it a priori} what
are the relevant degrees of freedom which are related to the glass
transition. Thus, $N_{corr,4}$ and $S_c$ have to be normalized to
a ``number of relevant degrees of freedom'' (or {\it beads}) per
unit volume. A procedure to define these beads has been proposed by
Stevenson and Wolynes \cite{SW05}; still the physical meaning of
these objects is unclear. On the contrary, $\sigma_{_{CRR}}$
represents the number of states accessible to a CRR
and is independent of the number of relevant degrees of freedom
inside this region. In fact, in Eq.~(\ref{Scorr}) the dependence
on the beads density is cancelled by taking the ratio $S_c/\D
C_p$.

\section{Data analysis}

For each analyzed material, we collected original and literature
data for the relaxation time $\t_\a(T)$, the stretching exponent
$\b(T)$, the configurational entropy $S_c(T)$ and heat capacity
$\D C_p(T)$ as functions of temperature at constant pressure. 
From these quantities we compute $N_{corr,4}$ and
$\s_{CRR}$ according to (\ref{Ncorr4-der}) and (\ref{Scorr}),
following the procedures already detailed in \cite{Da07}. Some
remarks on the data analysis are worth at this point: \\
1. $S_c$ and $\D C_p$ are not directly accessible by 
experiments. Following an approximation commonly adopted in literature, we estimated 
these quantities from the excess entropy of the supercooled liquid 
with respect to the crystal (we used the fits reported in the literature
or original data obtained for this work).
This approximation gave rise to several criticisms: for instance,
the difference of anharmonicity between the crystal and the supercooled liquid could 
reflect in a vibrational contribution to the excess entropy, see \eg~\cite{JOH00}. 
One should also consider the contribution of the 
secondary processes, when present~\cite{CANG}. 
All these discrepancies could induce an overestimation of $S_c$
up to $50\%$. Nevertheless, as the excess entropy
is believed to be often proportional to $S_c$, we used it as an estimation, 
as previously done in the literature. \\
2. The determination of the stretching parameter $\b$ 
is also a possible source of error. Sometimes it has been done
in the literature by fitting the whole correlation function and
subtracting the contribution of additional fast processes. 
These procedures are often model dependent, and can induce an error up to $10\%$ 
in the value of $\beta$. Indeed, in some papers, the $\beta$ of dielectric structural relaxation 
has been underestimated by using a fitting procedure on $\log-\log$ scale, that emphasized the 
high frequency tail of the loss peak. For these cases, $\beta$ has been estimated here
by using a fitting procedure mainly sensitive to the region of the full width at half maximum.\\
3. We fitted literature data for $\t_\a(T)$ by a Vogel-Fulcher-Tamman (VFT)
law, $\t_\a(T) = \t_0 e^{\frac{B}{T-T_0}}$, in a temperature range around $T_g$ where
reliable data are available. The VFT
parameters are reported in table~\ref{tabella1}.
For most of the systems, we found,
as already observed in the literature \cite{HAN97,RICH98}, that only one VFT law was not able to 
fit the data in the whole temperature range, and that only in a region close to $T_g$ the
Adam-Gibbs relation was satisfied: in these cases only data belonging to this region 
were reported.
In some cases we had not access to the raw
data but only to the VFT fit reported in the literature.
Note that the characteristic time $\t_0$ obtained from the VFT fit
is not the $T\to\io$ limit of $\t_\a(T)$ as the fit is
performed in an interval close to $T_g$ where data are available.
This is consistent since we are interested in the "effective"
physical attempt frequency close to $T_g$.
The derivative $d\ln \tau_\a/d\ln T$
was obtained from the VFT fit when it is reliable
(see \cite{HAN97} for details).
In the cases where VFT gave a poor fit, we directly derived the data 
(if dense enough) or we used a polynomial interpolation. \\
4. We do not report here a systematic analysis of experimental errors;
correspondingly error bars are not reported in the figures. Errors
are, generically, of the order of 10$\%$ on all quantities, mainly
coming from the difficulty in the determination of $\b$ and $S_c$ as discussed above.
For some materials
different datasets were available: we checked that the analysis
using different sources gives consistent results within these errors.
In tables~\ref{tabella1},\ref{tabella2},\ref{tabella3}
we report the list of the investigated
materials and the sources of the data.

\begin{figure}[t]
\centerline{\includegraphics[width=8cm]{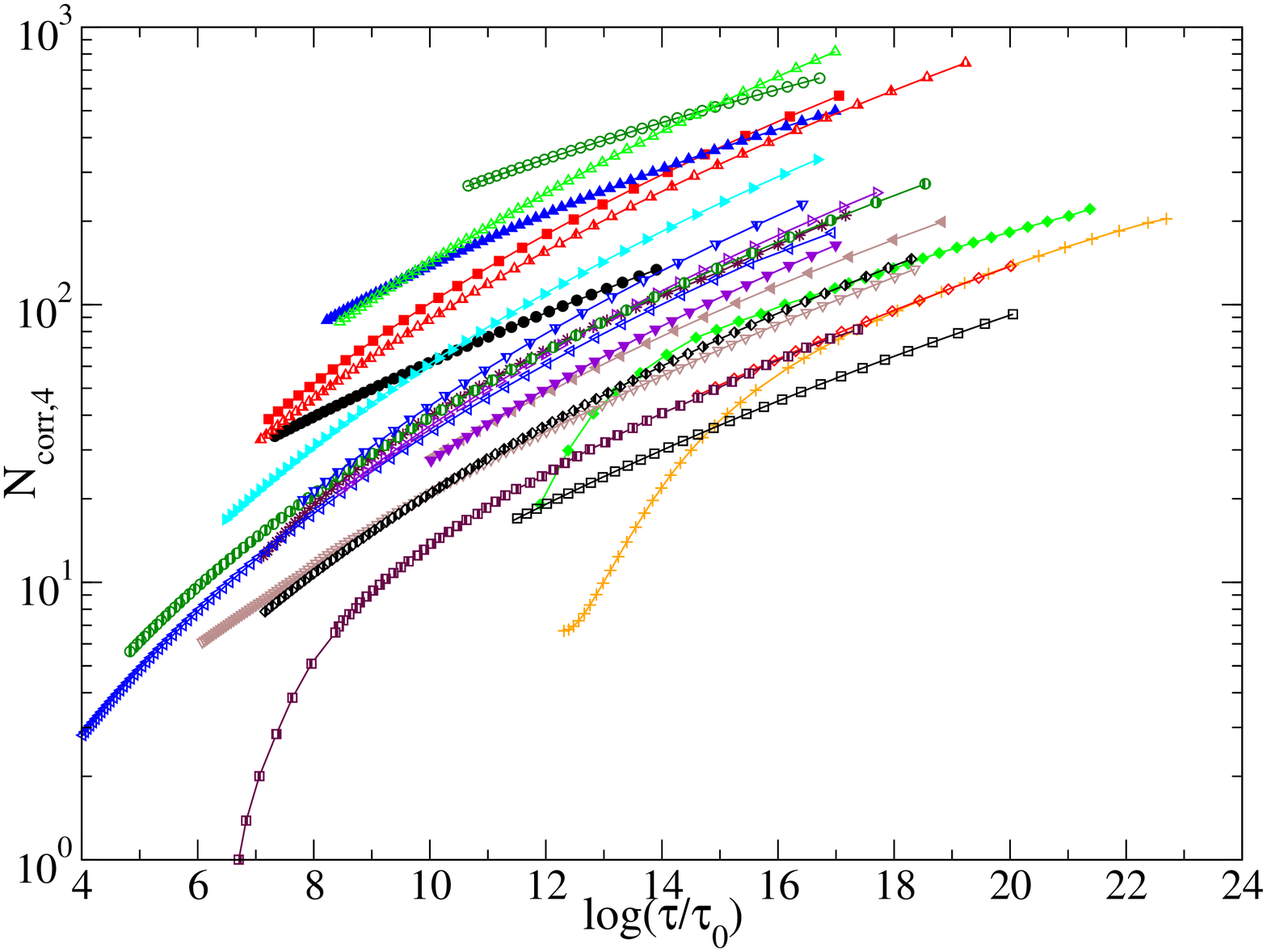}}
\vskip-10pt
\centerline{\includegraphics[width=8cm]{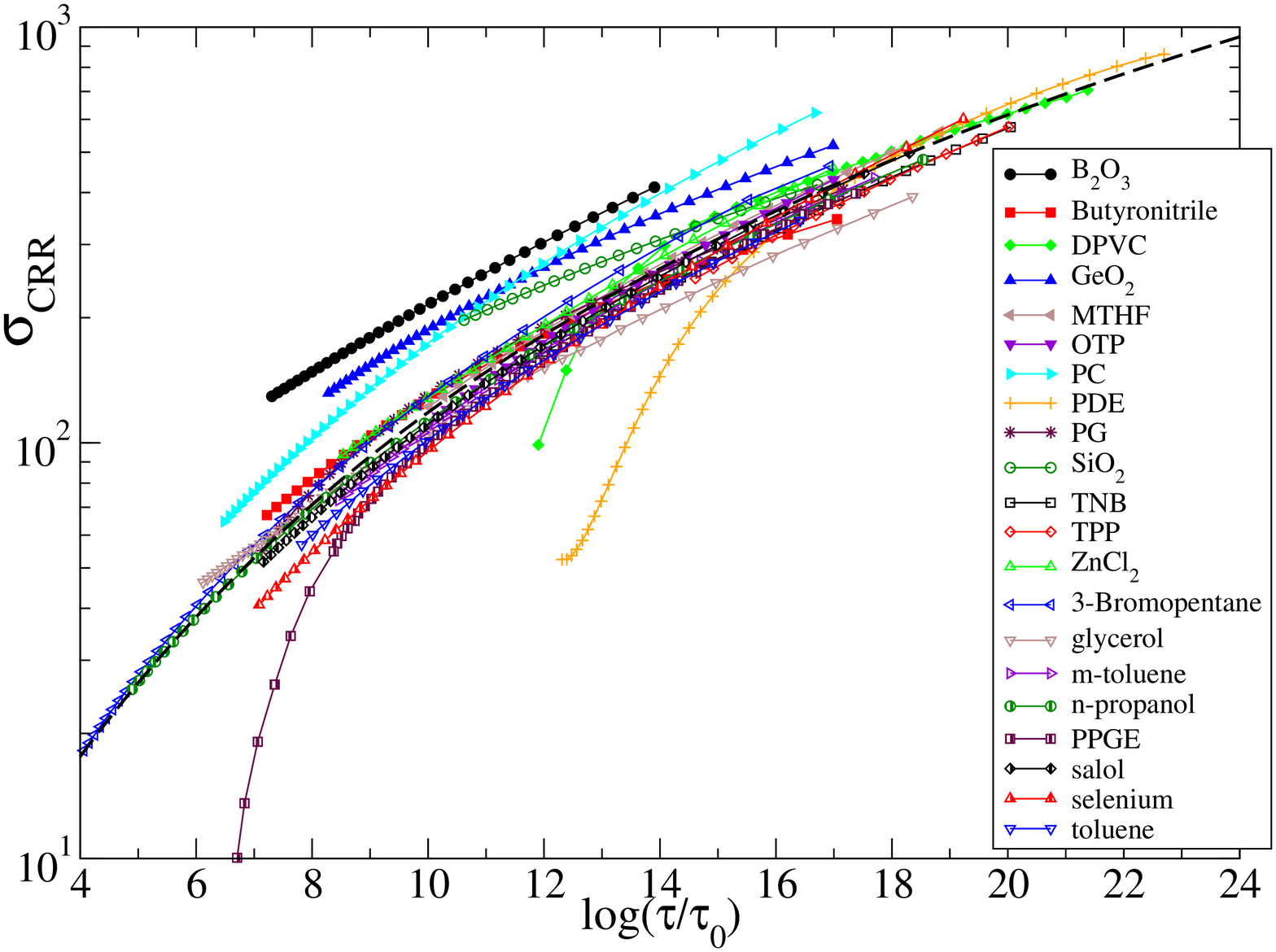}}
\caption{(Top) $N_{corr,4}$ as a function of $\log(\t_\a/\t_0)$
and (bottom) $\sigma_{_{CRR}}$ as a function of $\log(\t_\a/\t_0)$
for the materials listed in table~\ref{tabella1}.
In the lower panel the spreading of the
curves is, at fixed $\log(\t_\a/\t_0)$, about a factor of 2, while
in the upper panel is about a factor of 20. The dashed line is
$\log(\t_a/\t_0) = (\sigma/\sigma_o)^\psi + z
\ln(\sigma/\sigma_o) + \ln A$ with $A=0.65$, $\sigma_o=2.86$,
$z=1.075$, and $\psi=0.5$. \label{figNcorr}}
\end{figure}

To check the robustness of our results, we measured the
dielectric permittivity $\varepsilon(\omega)$, related to the dipole-dipole correlation,
of four different substances: 
OTP (see also \cite{Da07}), 
DPVC and PPGE, obtained from Sigma-Aldrich, and PDE that was kindly provided by 
Dr.~Marian Paluch, University of Katowice, Poland.
Frequency was spanned in the range 0.1 mHz-3 GHz using a combination of 
Novocontrol Alpha Analyser (up to 10 MHz) and Agilent Network Analyser (up to 3 GHz). 
A fine temperature scan gives direct access to the
temperature derivative $\chi_T(\omega) = \frac{d\e(\omega)}{dT}$,
from which we can determine $N_{corr,4}$ by
using Eq.~(\ref{Ncorr4-max}), where time is replaced by frequency.
The procedure we followed and the results are identical to the ones reported
in section III of \cite{Da07}.
Further experimental details will be given in~\cite{inprep}.
In figure~\ref{figconfronto} we report the values of $N_{corr,4}$
for these four substances determined by using
Eq.~(\ref{Ncorr4-max}) and (\ref{Ncorr4-der}): the close agreement
between the two estimates is a check of the validity of the
approximations discussed above. Then, for the rest of our
analysis, we will use Eq.~(\ref{Ncorr4-der})
to estimate $N_{corr,4}$.

\section{Results}

In figure~\ref{figNcorr}, top panel, we report the parametric plot
of $N_{corr,4}(T)$ versus $\tau_\alpha(T)$ for the 21
materials listed in table~\ref{tabella1}. Our results closely
agree with the ones already reported in \cite{Be06,Da07}: we
observe an increase of $N_{corr,4}$ on lowering the temperature,
consistently with the prediction of a growing cooperativity in
supercooled liquids. Note that we added to the plot some materials
that were not analyzed in \cite{Da07}; this is because the liquids
studied in \cite{Da07} all have similar values of $S_c(T_g)$ and
of VFT parameters, which then give similar values of $N_{corr,4}$
around $T_g$, see table~\ref{tabella1}. We added to the plot
materials like selenium, TPP and TNB whose configurational entropy
at $T_g$ is very different from the one of glycerol and other
molecular glass formers. For this reason the spreading of data in
our figure is more marked than in~\cite{Da07}, about a decade
at fixed $\log(\t_\a/\t_0)$.
Some of these systems are expected to severely test 
any kind of correlation between thermodynamic and dynamic properties, as 
they are also known as ``bad actors'' concerning the link between kinetic 
and themodynamic fragility \cite{WAN06}. 

In the bottom panel of figure~\ref{figNcorr} we show that the data
collapse much better on a universal curve when $\sigma_{_{CRR}}$
is plotted instead of $N_{corr,4}$. This result suggests that
$\t_\a(T)$ is given by a material-dependent characteristic attempt
rate $\t_0$ times a {\it universal function} of the thermodynamic
barrier $\sigma_{_{CRR}}(T)$, \ie \beq\label{univ} \log
[\t_\a(T)/\t_0] = \FF[\sigma_{_{CRR}}(T)] \ , \eeq with
$\FF[\sigma]$ an almost universal function.

\begin{figure}[t]
\centerline{\includegraphics[width=8cm]{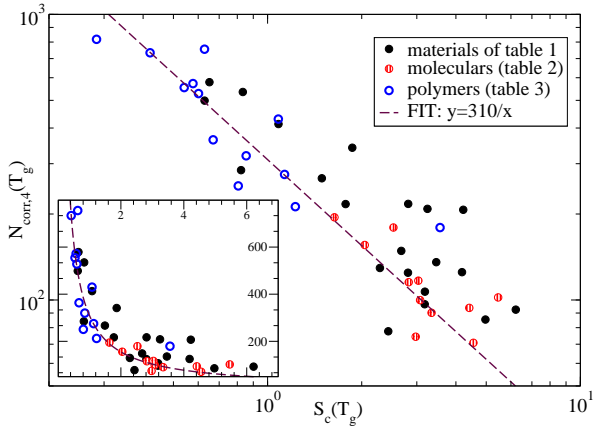}}
\caption{Plot of $N_{corr,4}(T_g)$ as a function of $S_c(T_g)$ (in
units of $R$) for the 45 materials listed in tables
\ref{tabella1},\ref{tabella2},\ref{tabella3}. The dashed line is a
fit to the inverse correlation predicted by Eq.~(\ref{correTg}).
In the inset the same plot is reported in linear scale.
\label{figcorreTg}}
\end{figure}

To give more robustness to this result one would like to add much
more materials to the plot. Unfortunately, for most materials
calorimetric data and/or the stretching exponent are reported only
at $T_g$, and the previous analysis is not possible. We thus
resort to a weaker test: for most materials $\t_0 \sim 10^{-15}$ s,
therefore the glass transition temperature $T_g$ is roughly
defined by $\log[\t_\a(T_g)/\t_0]=17$. If Eq.~(\ref{univ}) holds,
we expect
\beq\label{correTg} \sigma_{_{CRR}}(T_g) = \frac
{S_c(T_g)}{K_B} N_{corr,4}(T_g) \sim \FF^{-1}[17]  \ , \eeq
which implies an inverse correlation between $S_c$ and
$N_{corr,4}$ at $T_g$. These quantities can be easily computed
from the values of $S_c(T_g)$, $\D C_p(T_g)$, $\b(T_g)$ and
fragility $m = \frac{d\log\t_\a}{d\log T} |_{T=T_g} $, that are
reported in the literature for many more materials, see
tables~\ref{tabella2} and \ref{tabella3} for their list and the
references. We can then test the prediction (\ref{correTg}) on a
larger set of 45 materials. In figure~\ref{figcorreTg} we plot
$N_{corr,4}(T_g)$ as a function of $S_c(T_g)$; the plot is
compatible with an inverse correlation of these two quantities.

Note that from Eqs.~(\ref{correTg}), (\ref{Ncorr4-der}) it
follows easily that $S_c(T_g) \b^2 m^2 /\D C_p(T_g)=
\text{const.}$ Using the well-known relation $m/17 \sim \D
C_p(T_g)/S_c(T_g)$, that follows from the Adam-Gibbs relation
(\ref{AG}), we obtain $\b^2 m = \text{const.}$ The latter relation
has already been derived in \cite{LUB03} in a different way and
seems to be well verified in real materials, which is a nice
consistency check of our results.

\begin{figure}[t]
\centerline{\includegraphics[width=8cm]{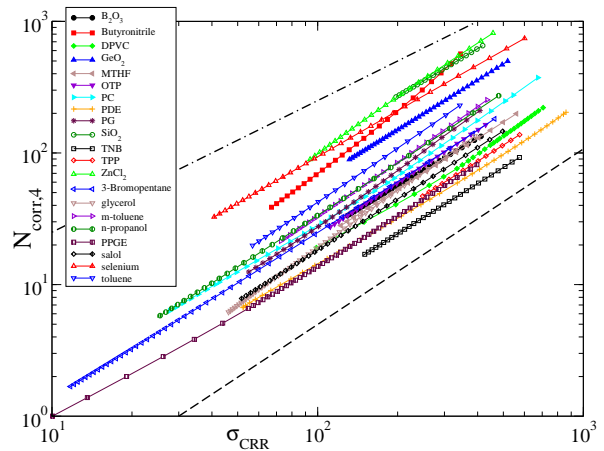}} \caption{
Parametric plot of $N_{corr,4}(T)$ as a function of
$\sigma_{_{CRR}}(T)$ for the materials in table~\ref{tabella1}.
The dashed line has slope 4/3 and the dot-dashed line has
slope~1.\label{figSN}}
\end{figure}

\section{Exponents}

Having stated our main experimental results, Eqs.~(\ref{univ}) and
(\ref{correTg}), we can turn to a more detailed comparison with
RFOT. First we note that the experimentally observed
increase of $\sigma_{_{CRR}}(T)$ on lowering the temperature is
not compatible with AG theory, while it fits well into the RFOT
scenario. Then we would like to extract the values of the RFOT
exponents from the experimental data: we anticipate that this is a
very difficult task given the large experimental errors already discussed 
and the probable importance of preasymptotic effects. Still we observe that,
identifying $N_{corr,4} \propto \rho \tilde \xi^d$ and using
Eqs.~(\ref{xiRFOT}), (\ref{deltaF}), RFOT predicts the relation
\beq\label{expo1} \sigma_{_{CRR}}(T) \sim
S_c(T)^{-\frac{\th}{d-\th}} \sim \tilde \xi^{\th} \sim
N_{corr,4}^{\th/d} \ . \eeq
In figure~\ref{figSN} we plot $N_{corr,4}$ as a function of
$\sigma_{_{CRR}}$ for the materials of table~\ref{tabella1}. The
power-law relation predicted by RFOT is very well verified, and
the resulting values for the exponent $\th$ are in the range $2.2
\div 2.5$, see table~\ref{tabella1}. Note that RFOT usually
assumes $\th =1.5$~\cite{LUB03}, while istantonic calculations
give $\th=2$~\cite{Fr05,DSW05}. This result is then quite
puzzling; note however that the relation $N_{corr,4} \sim \tilde
\xi^d$ is not well established \cite{Da07} and this could affect
the result for $\th$.

We can also try to estimate the exponent $\psi$ that relates the
relaxation time to $\sigma_{_{CRR}}$, $\t_\a \sim \t_0
e^{\sigma_{_{CRR}}^\psi}$. To account for preasymptotic effects,
we choose a specific form $\FF[\sigma] = (\sigma/\sigma_o)^\psi +
z \ln(\s/\s_0) + \ln A$~\cite{Da07} in Eq.~(\ref{univ}) to fit the
curves in the bottom panel of figure~\ref{figNcorr}. We obtain as
a best fit a value $\psi \sim 0.4 \div 0.5$; note however that, by
changing the other parameters, one can obtain quite good fits for
a range of values of $\psi = 0.3 \div 1.5$.

From $\log (\t_\a/\t_0) \sim \sigma_{_{CRR}}^\psi$ and
Eq.~(\ref{expo1}) one obtains
\beq \log \frac{\t_\a(T)}{\t_0} \propto S_c(T)^{-\frac{\th
\psi}{d-\th}} \ , \eeq
which is consistent with the Adam-Gibbs relation (\ref{AG}) only
if $\frac{\th \psi}{d-\th} = 1$. Assuming this and $\psi \sim 0.4
\div 0.5$, we get $\th \sim 2 \div 2.15$ which is consistent with
the values reported in table~\ref{tabella1}. The coincidence of
these two estimates of $\th$ might support the robustness of the
indication that its value should be close to or slightly larger
than 2.

The result for $\psi \sim 0.4 \div 0.5$ is less reliable. Still,
we obtain a strong indication that $\psi < 1$,
that implies that the relaxation time diverges {\it slower}
than the exponential of the thermodynamic barrier. This suggests
the existence of relaxation paths which are more efficient than
simple nucleation of a random state inside a CRR. A theoretical
description of such processes is still lacking.

\section{Conclusions}

We collected original and literature data on a set of 45
glass-forming materials and, exploiting the methods of
\cite{Be06,Da07}, we show that {\it i)}  the size of the CRR is
inversely correlated with the configurational entropy per unit
volume at $T=T_g$, Eq.~(\ref{correTg}) and {\it ii)} more
generally that the relaxation time seems to be a universal
function of the configurational entropy of a CRR,
Eq.~(\ref{univ}), in a wide range of temperatures around $T_g$. We
compared these results with theoretical predictions and found good
agreement with the RFOT scenario. We also gave an estimate of the
exponents $\th$ and $\psi$ of RFOT and found unexpected but still
reasonable values of these exponents.

We wish to stress again that in our analysis we made the strong
hypothesis that $N_{corr,4}$, as defined in \cite{Da07}, is
representative of the size of a CRR, \ie that $N_{corr,4} \propto
\r \tilde \xi^d$. A different scaling would not only change our
estimates for the RFOT exponents, but change the definition of
$\sigma_{_{CRR}}$ and ultimately destroy the scaling of the curves
in the bottom panel of figure~\ref{figNcorr}. The consistency of
our results seem to support the validity of this assumption, see
however~\cite{BerthierJCP, Da07} for a more detailed discussion.

\begin{acknowledgments}
We wish to thank Giulio Biroli for many useful discussions and in
particular for suggesting the plot in figure~\ref{figSN}, and the
authors of \cite{Da07} for sending us their data prior to
publication.
\end{acknowledgments}


\begin{table*}
\centering
\caption{
Summary of data at $T_g$ and references. When more databases were present, we compared the data
and verified the consistency. $[\bullet]$ indicates original data from this
work. Most of the dynamic data come from dielectric relaxation experiments, 
except for GeO$_2$ and SiO$_2$ (viscosity was used instead of $\tau$), for B$_2$O$_3$ and 
ZnCl$_2$ (photon correlation spectroscopy), 
selenium (mechanical relaxation).
Temperatures are in K, times in s, entropy and specific heat in units of R.
Note: PC=Propylene Carbonate, TNB=TriNaphthyl-Benzene, OTP=O-Terphenyl,
PDE=Phenolphtaleindimethyleteher, DPVC=Diphenyl-vinylidene Carbonate,
PPGE=Polyphenylglycidylether, PG=Propylene Glycol, MTHF=2-Methyltetrahydrofuran, TPP=Triphenylphosphite
\label{tabella1}}
\begin{tabular*}{\hsize}{@{\extracolsep{4.5pt}} ccccccccccccc}
\hline
Name & $T_g$ & $\log\tau_0$ & $B$ & $T_0$  & $S_c$ & $\D C_P$ & $\b$ & $m$ & $N_{corr,4}$ & $\theta$ & Ref. dyn. & Ref. calor. \\
\hline
PC  & 155 & -14.8 & 467 & 128 & 1.86 & 9.10 & 0.7 & 94 & 341 & 2.2 & \cite{SCHN99,STI96} & \cite{MOY00,NGAI99} \\
TNB  & 344 & -18.0 & 1620 & 264  & 6.22 & 18.1 & 0.56 & 86 & 92 & 2.4 & \cite{RICH03,ZHU86} & \cite{LUB03,MOY00,MIS00,MAG67,TSU96,PRI80}\\
OTP & 244 & -14.5 & 684 & 202  & 2.67 & 13.8 & 0.55 & 97 & 148 & 2.3 & \cite{HAN97,RICH98,RICH05,STI95},$[\bullet]$& \cite{RICH98,NGAI99,CHA72} \\
PDE & 294 & -20.7 & 1793 & 215 & 4.22 & 13.2 & 0.73 & 84 & 207 & 2.4 & \cite{STI95,CAS03a,CAS03b},$[\bullet]$ & $[\bullet]$ \\
DPVC & 251 & -19.0 & 1243 & 192 & 3.24 & 12.3 & 0.67 & 89 & 208 & 2.2 & \cite{CAP07},$[\bullet]$ & $[\bullet]$ \\
PPGE & 258 & -15.1 & 708 & 217  & 4.97 & 23.0 & 0.49 & 107 & 85 & 2.3 & \cite{COR02},$[\bullet]$ & \cite{COR03} \\
m-toluidine & 185 & -14.9 & 519 & 154 & 1.77 & 11.2 & 0.57 & 102 & 217 & 2.2 & \cite{MAN05,NIS07} & \cite{ALB99} \\
PG & 168 & -12.9 & 708 & 120  & 2.28 & 8.06 & 0.72 & 53 & 129 & 2.1 & \cite{LEO99} & \cite{ANG82,ANG97,KAU48,PAR99} \\
GeO$_2$ & 816 & -13.8 & 9732 & 199 & 1.08 & 0.75 & 1 & 21 & 413 & 2.4 & \cite{SIP01,BOH93,MAR01} & \cite{WAN06,MAR01} \\
SiO$_2$ & 1452 & -13.8 & 14562 & 530  & 0.65 & 0.37 & 0.7 & 25 & 579 & 2.5 & \cite{SIP01,BOH93,MAR01}  & \cite{SIP01} \\
ZnCl$_2$ & 385 & -12.8 & 1647 & 274 & 0.63 & 1.89 & 0.71 & 51 & 499 & 2.2 & \cite{PAV97} & \cite{ANG97,ANG77} \\
3Bromo-pentane & 108 & -12.9 & 374 & 83  & 2.81 & 9.09 & 0.62 & 64 & 124 & 2.3 & \cite{RICH98,NGAI98a} & \cite{RICH98} \\
MTHF & 90 & -17.3 & 406 & 69  & 2.81 & 8.91 & 0.62 & 83 & 217 & 2.3 & \cite{RICH98,NGAI98a} & \cite{RICH98} \\
n-propanol & 100 & -10.5 & 386 & 70  & 2.43 & 6.09 & 0.62 & 41 & 78 & 2.3 & \cite{RICH98,NGAI98a} & \cite{RICH98} \\
Salol & 221 & -15.9 & 823 & 175 & 3.46 & 13.2 & 0.58 & 86 & 135 & 2.3 & \cite{RICH98,NGAI98a} & \cite{RICH98} \\
Butyronitrile & 95 & -11.9 & 326 & 72  & 0.82 & 4.84 & 0.77 & 56 & 285 & 1.8 & \cite{ITO06} & \cite{HIK88} \\
TPP & 204 & -17.5 & 704 & 169  & 4.18 & 18.2 & 0.51 & 111 & 125 & 2.3 & \cite{NGAI99,SCHI96} & \cite{NGAI99} \\
Selenium & 309 & -15.5 & 1077 & 248  & 0.83 & 1.83 & 0.42 & 88 & 535 & 2.6 & \cite{BOH93b,ROL99} & \cite{ANG97,CHA74} \\
Glycerol & 188 & -14.7 & 973 & 130  & 3.18 & 10.0 & 0.68 & 54 & 97 & 2.1 & \cite{NGAI98a,MEN92,SCHO93} & \cite{MOY00,NGAI99} \\
Toluene & 116 & -15.1 & 328 & 97  & 1.49 & 8.71 & 0.55 & 103 & 267 & 2.2 &  \cite{DOS97,NGAI04} & \cite{NGAI99} \\
B$_2$O$_3$ & 553 & -10.7 & 2540 & 353  & 3.18 & 3.25 & 0.62 & 35 & 107 & 2.5 & \cite{SID93,SID07} & \cite{JOH00,WAN06,PRI80,ANG97,KAU48} \\
\hline
\end{tabular*}
\end{table*}

\begin{table*}
\caption{Data at $T_g$ and references for the molecular liquids used in figure~\ref{figcorreTg}.
Same units as in table~\ref{tabella1}.
\label{tabella2}}
\begin{tabular*}{\hsize}{@{\extracolsep{20pt}} cccccccccccc}
\hline
Name & $T_g$ & $S_c$ & $\D C_p$ & $\beta$ & $m$ & $N_{corr,4}$ & Ref. dyn. & Ref. calor. \\
\hline
Sorbitol & 268 & 4.41 & 28.8 & 0.48 & 128 & 94 & \cite{NGAI98a,HEN02} & \cite{WAN06} \\
Ethylbenzene & 115 & 2.82 & 9.67 & 0.68 & 58 & 115 & \cite{WAN06} & \cite{YAM98} \\
Isopropyl Benzene & 126 & 2.52 & 10.2 & 0.56 & 90 & 179 & \cite{NIS07,NGAI98a} & \cite{WAN06,NGAI99} \\
Triphenylethene & 248 & 5.46 & 14.5 & 0.5 & 91 & 102 & \cite{WAN06,JAK05} & \cite{JOH00,WAN06} \\
Ethylene Glycol & 153 & 2.04 & 7.58 & 0.78 & 52 & 156 & \cite{WAN06,MUR97} & \cite{JOH00,WAN06} \\
Ethanol & 97 & 1.63 & 5.46 & 0.7 & 55 & 195 & \cite{WAN06} & \cite{JOH00,WAN06} \\
3-methylpentane & 77 & 3.07 & 8.66 & 0.62 & 56 & 100 & \cite{WAN06,NGAI98a} & \cite{JOH00,WAN06} \\
Diethyl phthalate & 180 & 3.03 & 15.3 & 0.64 & 78 & 117 & \cite{PAW03} & \cite{JOH00} \\
a-phenyl-cresol & 220 & 3.34 & 15.4 & 0.53 & 83 & 90 & \cite{NGAI98a,MUR95} & \cite{MUR95} \\
Glucose & 309 & 2.97 & 17.3 & 0.37 & 115 & 74 & \cite{WAN06,GANG} & \cite{WAN06,ANG97,KAU48,STE05} \\
Indometacin & 318 & 4.55 & 19.8 & 0.59 & 75 & 71 & \cite{WAN06,CAR06} & \cite{WAN06} \\
\hline
\end{tabular*}
\end{table*}

\begin{table*}
\caption{Data at $T_g$ and references for the polymeric liquids used in figure~\ref{figcorreTg}.
Same units as in table~\ref{tabella1}.
For some systems reported by \cite{CANG},
a particular procedure, introduced there, was used to subtract from the excess entropy 
the contribution of the secondary relaxation.
\label{tabella3}}
\begin{tabular*}{\hsize}{@{\extracolsep{27pt}} cccccccccccc}
\hline
Name & $T_g$ & $S_c$ & $\D C_p$ & $\beta$ & $m$ & $N_{corr,4}$ & Ref. dyn. & Ref. calor. \\
\hline
PVC & 354 & 0.28 & 2.33 & 0.27 & 191 & 818 & \cite{NGAI98a,ROL99b} & \cite{CANG,ROL99b} \\
PET & 342 & 1.08 & 9.36 & 0.48 & 156 & 430 & \cite{NGAI98a,ROL99b} & \cite{CANG,ROL99b} \\
a-PMMA & 378 & 0.58 & 3.61 & 0.37 & 145 & 572 & \cite{NGAI98a,ROL99b} & \cite{CANG,ROL99b} \\
PS & 373 & 0.60 & 3.40 & 0.35 & 143 & 528 & \cite{NGAI98a,ROL99b} & \cite{CANG,ROL99b} \\
PP & 270 & 0.63 & 2.44 & 0.37 & 137 & 755 & \cite{NGAI98a,ROL99b} & \cite{CANG,ROL99b} \\
PDMS & 146 & 0.42 & 3.07 & 0.56 & 100 & 734 & \cite{NGAI98a,ROL99b} & \cite{CANG,ROL99b} \\
PIsop & 200 & 0.80 & 3.71 & 0.47 & 77 & 251 & \cite{NGAI98a,ROL99b} & \cite{CANG,ROL99b} \\
PPO & 195 & 1.13 & 3.86 & 0.52 & 74 & 275 & \cite{NGAI98a,ROL99b} & \cite{ROL99b} \\
PIB & 200 & 3.56 & 2.56 & 0.55 & 46 & 179 & \cite{NGAI98a,ROL99b} & \cite{ROL99b} \\
PE & 237 & 0.67 & 1.26 & 0.55 & 46 & 364 & \cite{NGAI98a,ROL99b} & \cite{CANG,ROL99b} \\
PEN & 390 & 0.85 & 10.1 & 0.48 & 140 & 320 & \cite{CANG,NGAI98a} & \cite{WAN06} \\
PEEK & 419 & 0.54 & 10.4 & 0.32 & 280 & 554 & \cite{CANG,NGAI98a} & \cite{CANG} \\
Pcarb & 420 & 1.22 & 7.21 & 0.35 & 132 & 212 & \cite{CANG,NGAI98a} & \cite{CANG} \\
\hline
\end{tabular*}
\end{table*}


\begin{thebibliography}{99}


\bibitem{AG65} G.~Adam and J.~H.~Gibbs, J.~Chem.~Phys. { 43}, 139 (1965).

\bibitem{BB04}
J.~P.~Bouchaud and G.~Biroli, J.~Chem.~Phys. { 121}, 7347 (2004).

\bibitem{KTW87} T.R.Kirkpatrick, P.~Wolynes, Phys.Rev.B { 36}, 8552 (1987);
T.R.Kirkpatrick, D.~Thirumalai, P.~Wolynes, Phys.Rev.A { 40}, 1045 (1989).

\bibitem{XW01}
X.~Xia and P.~G.~Wolynes, Proc.~Nat.~Acad.~Sci. { 97}, 2990 (2000);
Phys.~Rev.~Lett { 86}, 5526 (2001).

\bibitem{LUB03}
V.~Lubchenko and P.~Wolynes, J.~Chem.~Phys { 119}, 9088 (2003);
Annu. Rev. Phys. Chem. { 58}, 235 (2007).

\bibitem{Be06} L.~Berthier {\it et al.}, Science { 310}, 1797 (2005).

\bibitem{Da07}
C.~Dalle-Ferrier {\it et al.}, arXiv.org:0706.1906 (2007).


\bibitem{Fr05} S.~Franz, J.~Stat.~Mech. (2005) P04001.

\bibitem{DSW05} M.~Dzero, J.~Schmalian and P.~G.~Wolynes,
Phys.~Rev.~B { 72}, 100201 (2005).

\bibitem{Pa99} G.~Parisi, J.~Phys.~Chem. B { 103}, 4128 (1999);
S.~Franz, C.~Donati, G.~Parisi and S.~C.~Glotzer, Philos.Mag.B { 79}, 1827 (1999);
C.~Bennemann, C.~Donati, J.~Baschnagel, S.~C.~Glotzer, Nature (London) { 399},
246 (1999).

\bibitem{BerthierJCP}
C.~Toninelli et al., Phys.Rev.E { 71}, 041505 (2005); L.Berthier et al., J.Chem.Phys. 126, 184503 and 184504 (2007).

\bibitem{staticC} A.~Montanari and G.~Semerjian, J.~Stat.~Phys. { 125}, 23 (2006);
A.Cavagna, T.S.Grigera, P.Verrocchio, Phys. Rev. Lett. { 98}, 187801 (2007);
S.Franz and A.Montanari, J. Phys. A: Math. Theor. { 40}, F251-F257 (2007).

\bibitem{Ed00} M.~D.~Ediger, Ann.~Rev.~Phys.~Chem. { 51}, 99 (2000).

\bibitem{Dixon90} P.~K.~Dixon et al., Phys.~Rev.~Lett. { 65}, 1108 (1990).

\bibitem{SW05} J.~D.~Stevenson and P.~G.~Wolynes, J.~Phys.~Chem.~B { 109},
  15093 (2005).

\bibitem{JOH00} G.P. Johari, J. Chem. Phys. { 112}, 8958 (2000).

\bibitem{CANG} D. Cangialosi, A. Alegria, J. Colmenero, Europhys. Lett. { 70}, 614 (2005).

\bibitem{HAN97} C. Hansen, F. Stickel, T. Berger, R. Richert, and E. W. Fischer, J. Chem. Phys. { 107}, 
1086 (1997).

\bibitem{RICH98} R.~Richert, C.~A.~Angell, J.~Chem.~Phys. { 108}, 9016 (1998).

\bibitem{inprep} S.Capaccioli, D.Prevosto, G.Ruocco, F.Zamponi, in preparation.

\bibitem{WAN06}  Li-Min Wang, C. Austen Angell, and Ranko Richert, J. Chem. Phys. { 125}, 074505 (2006).





\bibitem{SCHN99} U. Schneider, P. Lunkenheimer, R. Brand, and A. Loidl, Phys. Rev. E, 59, 6924 (1999).

\bibitem{STI96} F.~Stickel, E.~W.~Fischer, R.~Richert, J.~Chem.~Phys. { 104}, 2043 (1996).

\bibitem{MOY00} C.~T.~Moynihan, C.~A.~Angell, J.~Non-Cryst.~Solids { 274}, 131 (2000).

\bibitem{NGAI99} K.~L.~Ngai, O.~Yamamuro, J.~Chem.~Phys. { 111}, 10403 (1999).

\bibitem{RICH03} R. Richert, K. Duvvuri, and LT Duong, J. Chem. Phys. 118, 1828 (2003)

\bibitem{ZHU86} X.R. Zhu, C.H. Wang, J. Chem. Phys. 84, 6086 (1986).

\bibitem{MIS00} R.K. Mishra, K.S. Dubey, Journal of Thermal Analysis and Calorimetry, 62,  687 (2000)

\bibitem{MAG67} J.H. Magill, J.Chem.Phys. 47, 2802 (1967)

\bibitem{TSU96} I. Tsukushi, O. Yamamuro, T. Ohta, T. Matsuo, H. Nakano and Y. Shirota, J. Phys.: Condens. Matter 8, 245-255. (1996)

\bibitem{PRI80} V.P. Privalko, J. Phys. Chem., 84, 3307-3312 (1980)

\bibitem{RICH05} R.~Richert, J.~Chem.~Phys. { 123}, 154502 (2005).

\bibitem{STI95} F. Stickel, PhD Thesis, Mainz University, Shaker, Aachen, 1995

\bibitem{CHA72} S. S. Chang and A. B. Bestul, J. Chem. Phys. 56, 503 (1972)

\bibitem{CAS03a} R.~Casalini, K.~L.~Ngai, C.~M.~Roland, Phys.~Rev.~B { 68}, 014201 (2003).

\bibitem{CAS03b} R. Casalini, M. Paluch and C. M. Roland, J. Phys.: Condens. Matter 15, S859 (2003)

\bibitem{CAP07} S Capaccioli, K Kessairi, D Prevosto, M Lucchesi, P A Rolla, J. Phys.: Condens. Matter 19, 205133 (2007)

\bibitem{COR02} S. Corezzi, M. Beiner, H. Huth, and K. Schroter,S. Capaccioli and R. Casalini, D. Fioretto, E. Donth, J.Chem Phys., 117, 2435 (2002)

\bibitem{COR03} S. Corezzi, PhD thesis, Perugia 2003

\bibitem{MAN05} A. Mandanici, M. Cutroni, R. Richert, J.Chem Phys., 122, 084508 (2005)

\bibitem{NIS07} K. Niss, C. Dalle-Ferrier, G. Tarjus and Ch.Alba-Simionesco, J. Phys.: Condens. Matter 19, 076102 (2007)

\bibitem{ALB99} Ch.. Alba-Simionesco, J. Fan, C.A. Angell, J.Chem Phys., 110, 5262 (1999)

\bibitem{LEO99} C. Leon, K. L. Ngai, and C. M. Roland, J.Chem Phys., 110, 11585 (1999)

\bibitem{ANG82} C.A. Angell, D.L. Smith,  J. Phys. Chem. 86, 3045-3052 (1982)

\bibitem{ANG97} C.A. Angell, J. Res. Natl. Inst. Stand. Technol. 102, 171 (1997)

\bibitem{KAU48} W. Kauzmann, Chem. Rev. 43, 219 (1948)

\bibitem{PAR99} I. S. Park, K. Saruta and S. Kojima, J. Thermal Anal., 57, 687-693 (1999)

\bibitem{SIP01} A. Sipp, Y.Bottinga, P.Richet, J. Non-Cryst. Solids, 288, 166 (2001)

\bibitem{BOH93} B\"ohmer, K. L. Ngai, C. A. Angell, and D. J. Plazek, J. Chem. Phys. 99, 4201 (1993)

\bibitem{MAR01} L.-M. Martinez and C. A. Angell, Nature (London) 410, 663 (2001)

\bibitem{PAV97} E. A. Pavlatou, S. N. Yannopoulos, and G. N. Papatheodorou, G.Fytas, J. Phys. Chem. B, 101, 8748 (1997)

\bibitem{ANG77} C. A. Angell, E. Williams, K. J. Rao, and J. C. Tucker, J. Phys. Chem., 81, 238 (1977)

\bibitem{NGAI98a} K.L. Ngai, Physica A, 261, 36 (1998)

\bibitem{ITO06} N. Ito, K. Duvvuri, D. V. Matyushov, R.Richert, J. Chem. Phys. 125, 024504 (2006)

\bibitem{HIK88} H. Hikawa, M. Oguni, and H. Suga, J. Non-Cryst. Solids 101, 90 (1988)

\bibitem{SCHI96} B. Schiener, A. Loidl, R. V. Chamberlin, and R. B\"ohmer, J. Mol. Liq. 69, 243 (1996)

\bibitem{BOH93b} R. B\"ohmer and C.A. Angell, Phys.Rev.B, 48, 5857 (1993)

\bibitem{ROL99}  C.M. Roland, P.G. Santangelo, D. J. Plazek and K. M. Bernatz, J. Chem. Phys., 111, 9337 (1999)

\bibitem{CHA74} S. S. Chang and A. B. Bestul, J. Chem. Thermodyn. 6, 325 (1974)

\bibitem{MEN92} N. Menon, K. P. O'Brien, P.K. Dixon, L.Wu, S.R. Nagel, B.D. Williams and J.P. Carini, J. Non-Cryst. Solids, 141, 61 (1992)

\bibitem{SCHO93} A. Schonhals, F. Kremer, A. Hofmann, E.W. Fischer, E. Schlosser, Phys. Rev. Lett., 70, 3459 (1993)

\bibitem{DOS97} A. D\"oss, G. Hinze, B. Schiener, J. Hemberger, R. B\"ohmer, J. Chem. Phys. 107 (1997) 1740.

\bibitem{NGAI04} K.L. Ngai, S. Capaccioli, Phys. Rev. E 69, 031501 (2004)

\bibitem{SID93} D. L. Sidebottom, R. Bergman, L. B\"orjesson, and L. M. Torell, Phys. Rev. Lett. 71, 2260 (1993)

\bibitem{SID07} D. L. Sidebottom, B. V. Rodenburg, and J. R. Changstrom, Phys. Rev. B, 75, 132201 (2007)






\bibitem{HEN02} S. Hensel-Bielowka, M. Paluch, J. Ziolo, and C. M. Roland, J. Phys. Chem. B, 106, 12459 (2002)

\bibitem{YAM98} O. Yamamuro, I. Tsukushi, A. Lindqvist, S. Takahara, M. Ishikawa, and T. Matsuo, J. Phys. Chem. B 102, 1605 (1998)

\bibitem{JAK05} B Jakobsen, K. Niss and N.B. Olsen, J. Chem. Phys., 123, 234511 (2005)

\bibitem{MUR97} S.S.N. Murthy, J. Phys. Chem. B, 101, 60, 43 (1997)

\bibitem{PAW03} S. Pawlus, M. Paluch, M. Sekula, K. L. Ngai, S. J. Rzoska,1 and J. Ziolo, Phys. Rev. E, 68, 021503 (2003)

\bibitem{MUR95} S. S. N. Murthy, A. Paikaray, and N. Arya, 102, 8213 (1995)

\bibitem{GANG} Gangasharan, S.S.N. Murthy, J. Phys. Chem., 99, 12349 (1995); Gangasharan and S.S.N. Murthy, J. Chem. Phys. 99, 9865 (1993)

\bibitem{STE05} J.D. Stevenson, P.G. Wolynes, J. Phys. Chem. B, 109, 15093 (2005)

\bibitem{CAR06} L. Carpentier, R. Decressain, S. Desprez, and M. Descamps, J. Phys. Chem. B 110, 457 (2006)

\bibitem{ROL99b} C. M. Roland, P. G. Santangelo, and K. L. Ngai, J. Chem. Phys., 111, 5593 (1999)



\end{thebibliography}
\end{document}